\documentclass[]{aa}
\usepackage{graphicx}
\usepackage{comment}
\begin{document}
\title{Surface gravity of neutron stars and strange stars}
\author{M. Bejger\inst{1}
 and P. Haensel\inst{1,2}
}
 \institute{N. Copernicus Astronomical
Center, Polish
           Academy of Sciences, Bartycka 18, PL-00-716 Warszawa, Poland
\and LUTH  du CNRS, Observatoire de Paris, F-92195 Meudon Cedex, France\\
{\tt bejger@camk.edu.pl,  haensel@camk.edu.pl
 }}
\offprints{M. Bejger}
\date{Received 20 October 2003  / Accepted 28 February 2004 }
\abstract{ The upper bound on the value of the surface gravity,
$g_{\rm s}$, for neutron stars with equations of state respecting
$v_{\rm sound}\le c$, is derived. This bound is inversely
proportional to the  maximum allowable mass $M_{\rm max}$, and it
reads $g_{\rm s}\le 1.411 \times 10^{15}\;({M_{\rm max} /{
M}_\odot})^{-1}~{\rm cm~s^{-2}}$. It  implies an {\it absolute
upper bound} $7.4\times 10^{14}~{\rm cm~ s^{-2}}$ if one uses the
$2\sigma$ lower bound on the neutron mass measured recently in
4U1700-37, $1.9 M_{\odot}$. A correlation between $g_{\rm s}$ and
the compactness parameter $2GM/Rc^2$ for baryonic stars is
analyzed. The properties of $g_{\rm s}$  of strange quark stars
and its upper bounds are discussed using the scaling properties of
the strange-star models.
}
\titlerunning{Surface gravity of compact stars}
\authorrunning{P. Haensel and M. Bejger}
\maketitle
\section{Introduction}
\label{sect:introduction}
Gravitational acceleration on the stellar surface, usually called
the surface gravity, is an important parameter of the theory of
stellar atmospheres, in particular the atmospheres of neutron
stars (see, e.g., Zavlin \& Pavlov 2002). For  neutron stars,
$g_{\rm s}$ should be defined taking into account
 the space-time curvature. The value of $g_{\rm s}\sim {\rm few}\times
 10^{14}~{\rm cm~s^{-2}}$  is involved the relation between internal and
 surface temperature of neutron stars (Potekhin et al. 2003 and
 references therein).

The surface gravity of neutron stars is by many orders of
magnitude larger than for other stars; it is $\sim 10^5$ times
larger than for white dwarfs, and $10^8$ times stronger than the
gravity at the solar surface. As we show in the present paper, at
a given neutron-star mass the value of $g_{\rm s}$ depends very
strongly on the largely unknown equation of state (EOS) of dense
matter at supra-nuclear densities. It is therefore of interest to
derive upper bounds on $g_{\rm s}$. In Sect.\ \ref{sect:gs.bounds}
we derive such upper bounds on $g_{\rm s}$ resulting from the
condition of subluminality of the EOS (speed of sound not
exceeding $c$). We show how an upper bound on $g_{\rm s}$ can be
obtained from measured neutron star masses. In Sect.\
\ref{sect:gs.EOS} we analyze the EOS-dependence of  $g_{\rm s}$
for a set of 31 models of dense matter and compare the
maximal surface gravity, reached at the maximum allowable mass,
with subluminal upper bounds. In Sect.\ \ref{sect:gs.SS.scaling} we
consider surface gravity of strange quark stars. We derive a scaling
formula for $g_{\rm s}$ of strange stars and use it to relate the
maximum surface gravity for various models of such stars. Concluding
remarks are  presented in  Sect.\ \ref{sect:conclusion}.
\section{Bounds on surface gravity}
\label{sect:gs.bounds}
It is well known that (see, e.g., Shapiro \& Teukolsky 1983)
\begin{equation}
 g_{\rm s}=  { GM\over R^2\sqrt{1- 2GM/Rc^2}}~,
\label{eq:gs.MR}
\end{equation}
where $M$ is the gravitational mass of the star and $R$ is the stellar
circumferential radius, and effects of rotation are neglected. For
a ``standard neutron star'' with $M= 1.4\;M_\odot$ and $R=10$ km
one has $g_{\rm s}=2.43 \times 10^{14}~{\rm cm~s^{-2}}$. It is
therefore convenient to measure $g_{\rm s}$ in units of
$10^{14}~{\rm cm~s^{-2}}$, and to use $g_{\rm s,14}\equiv g_{\rm
s}/(10^{14}~{\rm cm~s^{-2}})$.

 Let us  introduce the dimensionless compactness parameter
  $x\equiv r_{\rm g}/R=2GM/Rc^2$ where $r_{\rm g}$ is the Schwarzschild
  radius. Then
\begin{equation}
 g_{\rm s,14}(x,M)= 15.21\;
 {x^2\over \sqrt{1- x}}\; {M_\odot\over M}~.
 \label{eq:gs.Mx}
 \end{equation}

 Generally, physical theories have to be causal. In the literature,
the condition
 of {\it causality} is usually replaced by the constraint  of {\it
 subluminality} of the EOS,
 $v_{\rm sound}=({\rm d}P/{\rm d}\rho)^{1/2}\le c$
 (see, e.g., Rhoades \& Ruffini 1974, Hartle
 1978, Kalogera \& Baym 1996, Keiser \& Polyzou 1996, 
Glendenning 1997, Koranda et al. 1997,
 Haensel et al.  1999), but strictly speaking, subluminality is not
 equivalent to causality. For instance, Bludman \& Ruderman (1968)
 constructed  causal and Lorenz-invariant EOSs with
 $v_{\rm sound}>c$. In practice, the difference seems to be
 unimportant; the counterexamples of Bludman \& Ruderman (1968)
 correspond actually to a highly excited medium
 (Bludman \& Ruderman 1970).
  Nevertheless, for the sake of rigor, we will
   call an EOS  respecting the condition
   $v_{\rm sound}\le c$  ``subluminal'', and an EOS violating
this condition will be called ``superluminal''.

 For subluminal  EOSs of dense matter, a strict upper bound on $x$
 is $x_{\rm max}=0.708$ (see, e.g.,  Haensel et al. 1999
 and references therein). Using this value we get, from Eq.\ (\ref{eq:gs.Mx}),
 an upper bound for a surface gravity of non-rotating neutron star of mass $M$,
\begin{equation}
v_{\rm sound}\le c ~~~~\Longrightarrow ~~~~ g_{\rm s,14}\le g^{\rm
SL}_{\rm max,14}= 14.11 \; {M_\odot\over M}~.
\label{eq:UB.M.sublum}
\end{equation}

The maximal mass $M_{\rm max}$ 
obtained from the realistic EOS should 
be higher than the largest measured neutron star mass, 
$M^{\rm max}_{\rm obs}$. Known rotation frequencies of neutron stars
are sufficiently small compared to the mass-shedding (Keplerian)
limit, therefore we can use the non-rotating approximation for
observed neutron stars. We thus have, in terms 
of $g_{\rm s}$, an upper bound based on the observations and subluminality 
conditions

\begin{equation}
g_{\rm s,14}<14.11 \; {M_\odot \over M^{\rm max}_{\rm obs}}~.
\label{eq:UB.Mobs.sublum}
\end{equation}
The higher $M^{\rm max}_{\rm obs}$, the stronger the constraints on the
maximum allowable $g_{\rm s}$ (the smaller the value of the upper bound). 
A precisely measured mass of the Hulse-Taylor
pulsar, $(1.4408\pm0.0006)~{\rm M}_\odot$ (Weisberg \& Taylor
2003), gives the upper bound $9.79\times 10^{14}~{\rm
cm~s^{-2}}$. According to Clark et al. (2002), neutron star in the
high-mass X-ray binary 4U1700-37 has, at the $ 2\sigma$ confidence
level, $M_{\rm obs}>1.9~{\rm M}_\odot$. This implies an upper
bound {\it lower than} $7.4\times 10^{14}~ {\rm cm~s^{-2}}$. It
should be mentioned, however, that the result of Clark et al.
(2002) has to used with care, because of large errors. Moreover,
because the compact object in 4U1700-37 is neither an X-ray pulsar
nor an X-ray burster, it could in principle be a low-mass black
hole.

The expression on the right-hand-side of (\ref{eq:UB.Mobs.sublum})
 deserves an additional comment. Apart from the numerical constant,
it is identical to the upper bound on the frequency of stable
rigid rotation of neutron stars with subluminal EOS (Koranda et
al. 1997). An alternative derivation of an approximate (but very
precise) upper bound on rotation frequency, relating it to the
maximum stellar compactness (surface redshift) for subluminal EOS,
was given by Haensel et al. (1999). In the latter work, the
starting point was the ``empirical formula'' for the maximum
rotation frequency. Its  upper bound was also obtained by
maximizing a function of $x=r_{\rm g}/R$.
\section{Equation of state of dense matter and surface gravity}
\label{sect:gs.EOS}
We have calculated $g_{\rm s}$ for neutron-star models based on a
set of 31 baryonic EOSs of dense matter.
 The EOSs were obtained under different assumptions regarding the
 composition of the matter at $\rho\ga 2\rho_0$, where 
$\rho_0 = 2.7\times  10^{14}~{\rm g~cm^{-3}}$ is the normal nuclear density.

The EOSs can be subdivided into several groups. Within one group
of models, the  matter consists of
 nucleons and leptons (Baldo et al. 1997,
Bombaci 1995,  Balberg \& Gal 1997,  Balberg et al. 1999, model I
of Bethe \& Johnson 1974 (usually called BJI), Pandharipande 1971,
Pandharipande \& Ravenhall 1989, Douchin \& Haensel 2001, Wiringa
et al. 1988, Walecka 1974, Haensel et al. 1980,
Akmal et al. 1998). Within the second group, the matter is assumed
to consist of nucleons, hyperons and leptons (Glendenning 1985,
Balberg \& Gal 1997, Weber et al. 1991). The third group involves
 an exotic high-density phase: de-confined quark matter mixed
 with baryonic matter (Glendenning 1997), pion condensate
(Muto \& Tatsumi 1990) and kaon condensate (Kubis 2001). Finally,
one EOS is the so called ``maximally-stiff-core'' (MSC) EOS. It consists
of the BJI  EOS below the baryon density $0.3~{\rm fm}^{-3}$,
matched continuously to the EOS with $v_{\rm sound}=c$ at higher density.

\begin{figure}
\resizebox{\hsize}{!}{\includegraphics[angle=0]{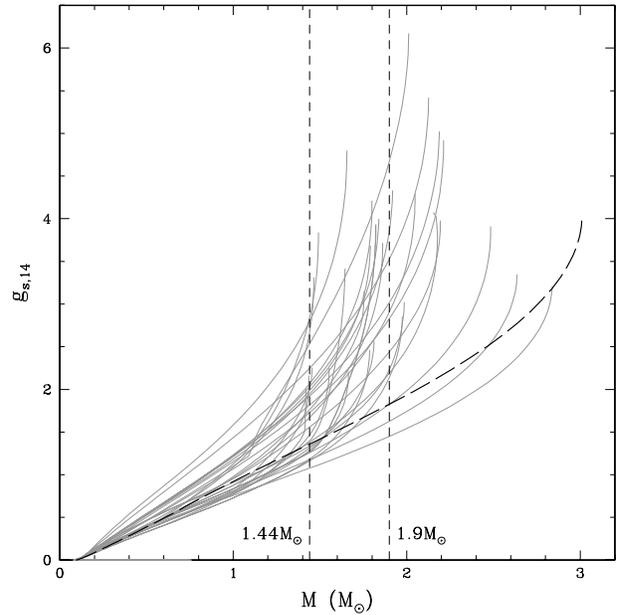}}
\caption{Surface gravity in the units of $10^{14}~{\rm cm~s^{-2}}$
against neutron-star mass for 31 EOSs of dense baryonic
matter. Only  stable configurations are shown, so that the curves
terminate at the maximum allowable mass. Thick dashed line:
maximally-stiff-core EOS. Remaining EOSs: thin lines. For
further explanation see the text. }
 \label{fig:gs.M}
\end{figure}
%
\begin{figure}
\resizebox{\hsize}{!}{\includegraphics[angle=0]{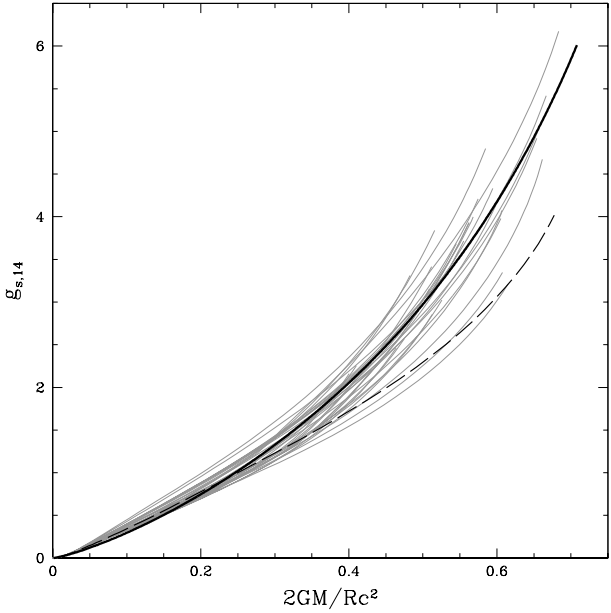}}
\caption{
Plots of $g_{\rm s,14}$ versus compactness parameter $2GM/Rc^2$. Notation as
in Fig.\ \ref{fig:gs.M}. Thick solid line represents approximate formula,
Eq.\ (\ref{eq:gs.x.fit}).}
 \label{fig:gs.comp}
\end{figure}

\begin{figure}
\resizebox{\hsize}{!}{\includegraphics[angle=0]{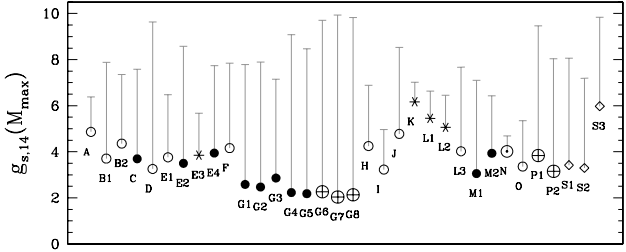}}
\caption{ Maximum values of surface gravity in units of
$10^{14}~{\rm cm~s^{-2}}$ (centers of the symbols at the bottom
end of the vertical segments) for 34 baryonic and quark EOSs of 
dense matter, and the upper-bounds at the maximum allowable 
mass for these EOSs obtained
 from Eq.\ (\ref{eq:UB.M.sublum}) (dashes at
 the upper end of the vertical  segments). Nucleonic EOS:  B1,B2 - Baldo
 et al. (1997);
D - Bombaci (1995); E1, E3 Balberg \& Gal (1997), Balberg et al.
(1999); C - model I of Bethe \& Johnson (1974); J - Pandharipande
(1971); F -  Pandharipande \& Ravenhall (1989); H - Douchin \&
Haensel (2001); L1, L2, L3 - Wiringa et al. (1988); 0 - Walecka
(1974); I - Haensel et al. (1980); A - Akmal et al.
(1998). Hyperonic EOSs: G1-G5 - Glendenning (1985); E2, E4
- Balberg \& Gal (1997), Balberg et al. (1999); M1, M2 -  Weber et
al. (1991). EOSs with exotic high-density phase: G6-G8 -
Glendenning (1997); K - Kubis (2001); P1, P2 - Muto \& Tatsumi (1990). 
Strange quark matter: S1, S2 -  Zdunik et al. (2000); S3 - Dey et al. (1998). 
N labels the MSC (maximally-stiff-core) EOS. For further 
explanations see the text.}
 \label{fig:gsmax.EOS}
\end{figure}

In Fig.\ \ref{fig:gs.M} we show the surface gravity versus stellar
(gravitational) mass for all 31 baryonic EOSs. The values
of $g_{\rm s}$ for a given  mass strongly depend on the EOS. For
$M=1.44~M_\odot$, $g_{\rm s,14}$ ranges from 1.1 to 2.8, while for
$M=1.9~M_\odot$ the predicted values are between 1.4 and 4.7. 

We considered also three EOSs of self-bound absolutely stable
strange quark matter, forming hypothetical strange stars (Zdunik
et al. 2000, Dey et al. 1998) - they will be studied  separately
in Sect.\ \ref{sect:gs.SS.scaling}.

Contrary to the significant scatter of the $g_{\rm s}(M)$ plots for
baryonic stars, $g_{\rm s}(x)$ is much less EOS-dependent: it can
be rather well (within better than 20\%) reproduced by the
approximate formula
\begin{equation}
g_{\rm s,14}\simeq {5x^{5/4}\over \sqrt{1-x}}~.
\label{eq:gs.x.fit}
\end{equation}
 Actually, the precision of this formula is much higher if we
exclude the superluminal EOSs and the unrealistically stiff ones
(too high incompressibility of nuclear matter at saturation) which
yield $M_{\rm max}\ga 2.5~{\rm M}_\odot$. Putting then the upper
bound $x_{\rm max}=0.708$ into (\ref{eq:gs.x.fit}) yields an
approximate ``realistic upper bound'' of $6\times 10^{14}~{\rm
cm~s^{-2}}$ on $g_{\rm s}$ of neutron stars.

The maximum value of $g_{\rm s}$ for stable stars with a given EOS
is reached at the maximum allowable mass. The values of $g_{\rm
s,max}$ for the selected EOSs are shown in Fig.\
\ref{fig:gsmax.EOS}. For comparison, we show also the
absolute upper bounds, Eq.\ (\ref{eq:UB.M.sublum}), at $M_{\rm
max}$. The dense matter models were divided into groups denoted
by specific symbols. Subluminal models involving only nucleons
($\circ$) give $g_{\rm s,max}\simeq (3-5)\times 10^{14}~{\rm
cm~s^{-2}}$. The stiffer the EOS, the closer $g_{\rm s,max}$ to
the subluminal upper bound $g^{\rm SL}_{\rm max}(M_{\rm max})$.

Subluminal hyperonic EOSs ($\bullet$) give $g_{\rm s,max}$ which
is typically lower than $4\times 10^{14}~{\rm cm~s^{-2}}$. For
these  EOSs, $g_{\rm s,max}$ can be as small as one-fifth  of the
upper bound $g^{\rm SL}_{\rm max}(M_{\rm max})$.

The subluminal EOSs with an exotic high-density phase ($\oplus$) have
relatively low $g_{\rm s,max}$. A phase transition softens the
EOS, lowering the radius. Simultaneously, however, the softening
leads to a decrease of $M_{\rm max}$. The latter effect dominates
over the former one. This strongly pushes up the upper bound
$g_{\rm s}(M_{\rm max})$. If the stellar interior consists mostly
of a mixed quark-baryon phase (EOSs G6, G7, G8 in Fig.\
\ref{fig:gsmax.EOS}), then  $g_{\rm s,max}\simeq 2\times
10^{14}~{\rm cm~s^{-2}}$, only one fifth of $g_{\rm max}^{\rm
SL}(M_{\rm max})$.

The  MSC EOS ($\odot$) yields $g_{\rm s,max}$ which is quite
close to $g_{\rm s,max}^{\rm SL}$. A similar situation occurs for
the EOSs that give $M_{\rm max}$ models with superluminal cores,
labeled $\star$. Their values of $g_{\rm s,max}$ range
from $4\times 10^{14}~{\rm cm~s^{-2}}$ to $6\times 10^{14}~{\rm cm~s^{-2}}$.
Note that even for those EOSs 
$g_{\rm s,max} < g_{\rm s,max}^{\rm SL}$. 
This is because the compactness parameter is always smaller than 
the upper bound for subluminal EOSs, $x(M_{\rm max}) < x_{\rm max} = 0.708$, 
as is shown in Fig. \ref{fig:gs.comp}.

\section{Surface gravity of strange stars}
\label{sect:gs.SS.scaling}
\begin{figure}
\resizebox{\hsize}{!}{\includegraphics[angle=0]{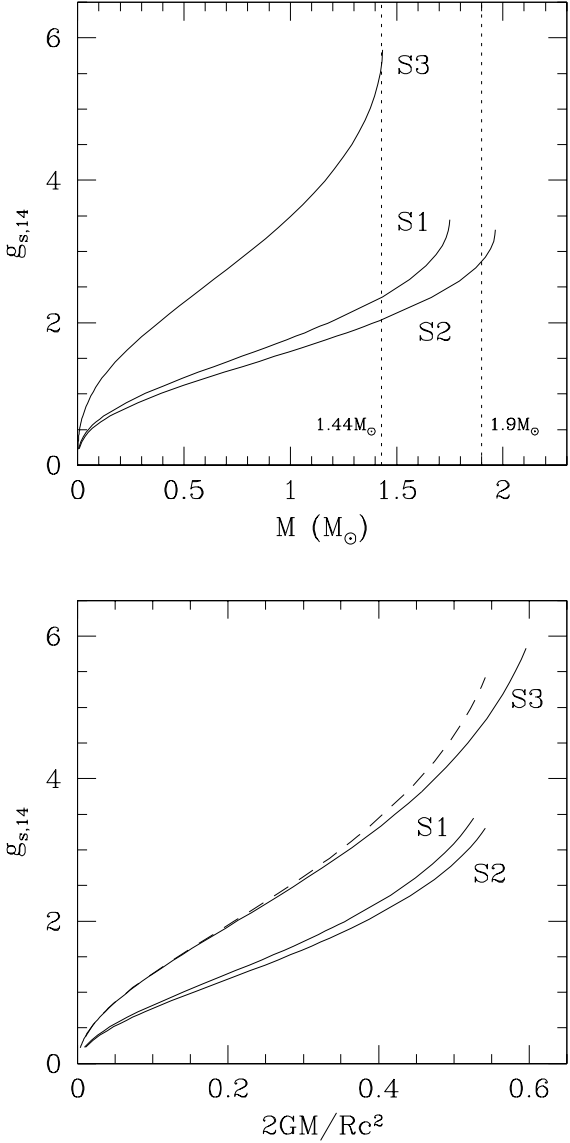}}
\caption{Surface gravity in units of $10^{14}~{\rm
cm~s^{-2}}$ versus strange star mass (upper panel)
and compactness parameter (lower panel), for three EOSs of strange
quark matter. Dashed line in the lower panel is obtained by transforming
(see text for details) the S2 EOS curve into S3 EOS curve using Eq.\
(\ref{eq:MR.SS.scaling}). }
 \label{fig:gs.SS}
\end{figure}

The $g_{\rm s}(M)$ dependence for strange stars, presented in the
upper panel of Fig.\ \ref{fig:gs.SS}, is very different from that
of baryonic stars, due to a different $R(M)$ dependence.
The model dependence  of $g_{\rm s}(M)$ for strange stars is very
strong: at $1.44~M_\odot$, the value of $g_{\rm s,14}$ ranges from
$2.0$ to $5.5$. However, this strong EOS dependence can be explained
in terms of the scaling properties of strange star models.

 EOSs of self-bound quark matter (strange
matter) are  derived from various different models of the quark
structure of hadrons. Despite differences between the
underlying models, the EOSs relevant for stable models of strange
stars  can be well represented (fitted) by a linear relation
between the pressure $P$ and the mass density $\rho$ (Zdunik
2000),
\begin{equation}
P=a c^2(\rho-\rho_{\rm s})~.
\label{eq:EOS.SQM}
\end{equation}
 The parameters of stellar
configuration, calculated for an EOS given by Eq.\
(\ref{eq:EOS.SQM}), are
 connected to those obtained for  an EOS with
 $\rho^\prime_{\rm s}\neq \rho_{\rm s}$
 by the   scaling relations (Witten 1984, Haensel et al. 1986,
Zdunik 2000). For example, points on the $M(R)$ curves are related by
\begin{equation}
M= \left({\rho^\prime_{\rm s}\over \rho_{\rm s}}\right)^{1\over 2}
M^\prime~,~~~~R= \left({\rho^\prime_{\rm s}\over \rho_{\rm
s}}\right)^{1\over 2}R^\prime~. \label{eq:MR.SS.scaling}
\end{equation}
Therefore, at a fixed $a$, the ratio $M/R$ (and therefore
$x=r_{\rm g}/R $) does not depend on $\rho_{\rm s}$,  and the
maximum surface gravity of strange stars scales as
\begin{equation}
g^\prime_{\rm s,max}=
\left({\rho^\prime_{\rm s}\over \rho_{\rm s}}\right)^{1\over 2}
g_{\rm s,max}~.
\label{eq:gs.SS.scaling}
\end{equation}

The values of $a$ for the EOSs S1, S2, and S3 range within
$0.30\le a \le 0.46$.  We have the exact ratio $g_{\rm s,max}({\rm
S3})/g_{\rm s,max}({\rm S2})=1.81$, while the scaling factor
$[\rho_{\rm s}({\rm S3})/\rho_{\rm s}({\rm S2})]^{1/2}=1.63$. The
large difference in maximum surface gravities can therefore be
accounted for by  scaling with respect to $\rho_{\rm s}$; in this
context, the $a$ dependence is sufficiently weak to be of
neglected in the simplest approach. As we see in the lower panel
in Fig.\ \ref{fig:gs.SS}, the precision of this scaling increases
with decreasing strange star compactness. The scaling becomes very
precise for $x<0.2$. The explanation of this behaviour is simple:
the density within such strange stars is nearly constant, so that
the dependence on $a$ is negligible.

\section{Discussion and conclusion}
\label{sect:conclusion}
We derived an upper bound on the surface gravity of static neutron
stars with subluminal EOSs.  Maximum $g_{\rm s}$ is reached at the
maximum allowable mass $M_{\rm max}$.  Even at the shortest observed
pulsar period of $1.56$ ms the  effect of rotation on the $M_{\rm
max}$ is very small, so that the static approximation is
justified. The upper bound on $g_{\rm s}$ is inversely
proportional to the highest measured neutron-star mass $M_{\rm
obs}^{\rm max}$ and is $7.4\times 10^{14}~{\rm cm~s^{-2}}$ if
$M_{\rm obs}^{\rm max}$ is replaced by $1.9~{\rm M}_\odot $, which
is the  $95\%$ confidence-level lower bound on mass of the compact
star in 4U1700-37 measured by Clark et al. (2002). However, this
upper bound on $g_{\rm s}$ should be taken as highly unreliable.
Firstly, Clark et al. (2002) used a specific model of the
companion star and of the binary to evaluate  the compact-star
mass. Secondly,  one cannot exclude, unfortunately,  that the
compact object in 4U1700-37 is actually a low-mass black hole and
not a neutron star (Clark et al. 2002 give arguments for and against 
the black-hole presence).

We have studied $g_{\rm s}$ for a set of 31 EOSs of
baryonic matter. The dependence of $g_{\rm s}$ on the stellar mass
$M$ is very sensitive to the EOS. On the contrary, the dependence
of $g_{\rm s}$ on the stellar compactness $r_{\rm g}/R$ has a
generic character for baryonic EOSs. The maximum surface gravity
$g_{\rm s,max}$ is sensitive to the EOS of dense matter and ranges
from about $2\times 10^{14}~{\rm cm~s^{-2}}$ to $5\times 10^{14}~{\rm
cm~s^{-2}}$ for subluminal baryonic EOSs.

The dependence of $g_{\rm s}$ on the mass and compactness of
strange stars is very different from that of baryonic stars.
However, the range of $g_{\rm s,max}=(3-6)\times 10^{14}~{\rm
cm~s^{-2}}$ is quite similar to baryonic stars.

We hope that the EOS-sensitive features of $g_{\rm s}$ will be 
useful in extracting information about the EOS of
dense matter, for instance, by combining the values  of $g_{\rm
s}$ obtained  fitting the thermal component of the observed photon
spectra with atmospheric models and the surface redshift measured
for the identified spectral lines.

\acknowledgement{This work was motivated by a question of  A.
Majczyna concerning the possible range of surface gravity in neutron
star atmospheres, which is gratefully acknowledged. We are grateful
to D.G. Yakovlev for the reading  of the manuscript  and for
helpful remarks. This work was partially supported by the KBN
grants no.  2P03D.019.24 and  5P03D.020.20. }


\begin{thebibliography}{} 

\bibitem[1998]{Akmal1998}
     Akmal, A., Pandharipande, V.R., \& Ravenhall, D.G. 1998,
     Phys. Rev. C, 58, 1804

\bibitem[1997]{Balberg97}
    Balberg, S., Gal, A. 1997,
    Nucl. Phys. A, 625, 435

\bibitem[1999]{BalbLC1999}
    Balberg, S., Lichtenstadt, I., \& Cook, G.B. 1999,
    ApJS, 121, 515

\bibitem[1997]{BBB97}
    Baldo, M., Bombaci, I., \& Burgio, G.F. 1997,
    A\&A, 328, 274

\bibitem[1974]{BetheJ1974}
    Bethe, H.A., \& Johnson, M.B. 1974,
    Nucl. Phys. A, 230, 1

\bibitem[1968]{BludmanRuderman1968}
    Bludman, S.A., \& Ruderman, M.A., 1968, Phys. Rev., 170, 1176

\bibitem[1970]{BludmanRuderman1970}
    Bludman, S.A., \& Ruderman, M.A., 1970, Phys. Rev. D, 1, 3243

\bibitem[1995]{Bombaci95}
    Bombaci, I. 1995, in Perspectives on Theoretical Nuclear Physics,
    ed. I. Bombaci, A. Bonaccorso, A. Fabrocini et al., 223

\bibitem[2002]{Clark2002}
Clark, J. S., Goodwin, S. P., Crowther, P. A., Kaper, L., 
Fairbairn, M., Langer, N., Brocksopp, C., 2002, A\&A, 392, 909

\bibitem[1998]{Dey1998}
Dey, M., Bombaci, I., Dey, J., Ray, S., \&  Samanta, B.C., 1998,
    Phys. Lett. B, 438, 123

\bibitem[2001]{DouchinHae2001}
Douchin, F., \& Haensel, P. 2001,
    A\&A, 380, 151

\bibitem[1985]{Glend1985}
    Glendenning N. K. 1985,
    ApJ, 293, 470
\bibitem[1997]{Glend1997BOOK}
    Glendenning, N.K. 1997, Compact Stars: Nuclear Physics,
    Particle Physics and General Relativity (Springer, New York)

\bibitem[1980]{HKP1980}
    Haensel, P., Kutschera, M., \& Pr{\'o}szy{\'n}ski, M. 1980,
    A\&A, 102, 299

\bibitem[1999]{HLZ1999}
    Haensel, P., Lasota, J. P., \& Zdunik, J. L., 1999,
    A\&A, 344, 151

\bibitem[2001]{HZS1986}
    Haensel P., Zdunik J. L., \& Schaefer R., 1986,
    A\&A, 160, 121

\bibitem[1978]{Hartle1978}
    Hartle, J.B., 1978, Phys. Repts., 46, 201

\bibitem[1996]{Kalogera1996}
    Kalogera, V., Baym, G., 1996, ApJ, 470, L61

\bibitem[1996]{Keiser1996}
    Keiser, B., Polyzou, W., 1996, Phys. Rev. C, 54, 2023

\bibitem[1997]{Koranda1997}
    Koranda, S., Stergioulas, N., Friedman, J.L., 1997,
    ApJ, 488, 799

\bibitem[2001]{Kubis2001}
    Kubis, S. 2001, Ph.D. Thesis, Jagiellonian University (unpublished)

\bibitem[1971]{Panda1971}
    Pandharipande, V.R. 1971,
    Nucl. Phys. A, 147, 641

\bibitem[1989]{PandhRav1989}
    Pandharipande, V.R., \& Ravenhall, D.G. 1989,
    in Proc. NATO Advanced Research Workshop on nuclear matter and
    heavy ion collisions, Les Houches, 1989, ed. M. Soyeur et al.
    (Plenum, New York, 1989), 103

\bibitem[2003]{PotekhinYakChabGned2003}
    Potekhin, A.Y., Yakovlev, D.G., \& Chabrier, G., 2003,
    ApJ, 594, 404

\bibitem[1974]{RhoadesRuff1974}
    Rhoades, C.E., \& Ruffini, R., 1974, Phys. Rev. Lett., 32, 324

\bibitem[1983]{ST1983}
Shapiro, S. L., Teukolsky, S. A., 1983, Black holes, white dwarfs, 
and neutron stars: The physics of compact objects, New York, Wiley-Interscience

\bibitem[1990]{tm90}
Muto T., Tatsumi T., 1990, Prog. Theoret. Phys. 83, 499

\bibitem[1974]{Walecka74}
    Walecka, J. D., 1974,
    Ann. Phys., 83, 491

\bibitem[1991]{WeberGW91}
    Weber, F., Glendenning, N.K., \& Weigel, M. K., 1991,
    ApJ, 373, 579

\bibitem[2003]{WeisbergTaylor2003}
    Weisberg,  J.M., Taylor, J.H., 2003,
    in: Radio Pulsars, edited by M. Bailes, Nice D.J., \& Thorsett S.E.,
    ASP Conference Series (to be published); e-print {\tt astro-ph/0211217}

\bibitem[1988]{WFF88}
    Wiringa, R.B.,  Fiks, V., \&  Fabrocini, A. 1988,
    Phys. Rev. C, 38, 1010

\bibitem[1984]{Witten1984}
Witten, E., 1984, Phys. Rev. D, 30, 272

\bibitem[2002]{ZavlinPavlov2002}
Zavlin, V.E., \&  Pavlov, G.G., 2002, in: Proceedings of the 270.
WE-Heraeus Seminar on Neutron Stars, Pulsars, and Supernova
Remnants. MPE Report 278. Edited by W. Becker, H. Lesch, and J.
Truemper. Garching bei Muenchen: Max-Planck-Institut fuer
extraterrestrische Physik, p. 263

\bibitem[2000]{Zdunik2000}
Zdunik, J.L., 2000,    A\&A, 359, 311


\end{thebibliography}
\end{document}